\documentclass{midl} 

\usepackage{soul}
\usepackage{mwe} 
\usepackage[utf8]{inputenc}
\usepackage{multirow}
\usepackage{microtype}
\usepackage{adjustbox}
\usepackage{caption}

\jmlrvolume{-- Under Review}
\jmlryear{2019}
\jmlrworkshop{Full Paper -- MIDL 2019 submission}
\editors{Under Review for MIDL 2019}

\title[Assessing Knee OA Severity with CNN attention-based end-to-end architectures]{Assessing Knee OA Severity with CNN attention-based end-to-end architectures}

\midlauthor{\Name{Marc Górriz\nametag{$^{1}$}} \Email{algayon2@gmail.com}\\
\Name{Joseph Antony\nametag{$^{2}$}} \Email{joseph.antony@dcu.ie}\\
\Name{Kevin McGuinness\nametag{$^{2}$}} \Email{kevin.mcguinness@dcu.ie}\\
\Name{Xavier Giró-i-Nieto\nametag{$^{1}$}} \Email{xavier.giro@upc.edu}\\
\Name{Noel E. O'Connor\nametag{$^{2}$}} \Email{noel.oconnor@dcu.ie}\\ \\
\addr $^{1}$ Universitat Politècnica de Catalunya (UPC), Barcelona, Catalonia/Spain \\
\addr $^{2}$ Insight Centre for Data Analytics, Dublin City University (DCU), Dublin, Ireland
}

\begin{document}

\maketitle

\begin{abstract}
This work proposes a novel end-to-end convolutional neural network (CNN) architecture to automatically quantify the severity of knee osteoarthritis (OA) using X-Ray images, which incorporates trainable attention modules acting as unsupervised fine-grained detectors of the region of interest (ROI). The proposed attention modules can be applied at different levels and scales across any CNN pipeline helping the network to learn relevant attention patterns over the most informative parts of the image at different resolutions. We test the proposed attention mechanism on existing state-of-the-art CNN architectures as our base models, achieving promising results on the benchmark knee OA datasets from the osteoarthritis initiative (OAI) and multicenter osteoarthritis study (MOST). All code from our experiments will be publicly available on the github repository: \url{https://github.com/marc-gorriz/KneeOA-CNNAttention}
\end{abstract}

\section{Introduction}
 Knee osteoarthritis (OA) is the most common articular disease and a leading cause of chronic disability \cite{heidari2011knee}, and mainly affects the elderly, obese, and those with a sedentary lifestyle. Degenerative processes of the articular cartilage as a result of excessive load on the joint, and the aging process, contributes to the natural breakdown of joint cartilage with joint space narrowing (JSN) and osteophytes \cite{KneeOA5}. Knee OA causes excruciating pain and often leads to joint arthroplasty in its severe stages. An early diagnosis is crucial for clinical treatment to be effective in curtailing progression and mitigating future disability \cite{KneeOA1} \cite{KneeOA2}. Despite the introduction of several imaging modalities such as MRI, OCT, and ultrasound for augmented OA diagnosis, X-Ray is still the method of choice in diagnosing knee OA, although clinical evidence also contributes.

Previous work has approached the challenge of automatically assessing knee OA severity as an image classification problem \cite{KneeOA2} \cite{KneeOA4} \cite{KneeOA6} using the Kellgren and Lawrence (KL) grading \cite{KL}. KL grading quantifies the degree of degeneration on a five-point scale (0 to 4): KL-0 (no OA changes), KL-1 (doubtful), KL-2 (early OA changes), KL-3 (moderate), and KL-4 (end-stage). Assessment is based on JSN, presence of osteophytes, sclerosis, and bone deformity. Most methods in the literature use a two step process to automatically quantify knee OA severity: 1) localization of knee joints; and 2) quantification of severity. Separate models for knee joint localization, either using hand-crafted features \cite{KneeOA4}, \cite{KneeOA6} or CNNs \cite{antony2} are not always highly accurate, affecting the subsequent quantification accuracy and adding extra complexity to the training process.

To overcome this problem, this work proposes a novel end-to-end architecture incorporating trainable attention modules that act as unsupervised fine-grained ROI detectors, which automatically localize knee joints without a separate localization step. The proposed attention modules can be applied at different levels and scales across an arbitrary CNN pipeline. This helps the network to learn attention patterns over the most informative parts of the image at different resolutions, achieving improvements in the quantification performance.

\section{Related work}
Much of the literature has proposed image classification-based solutions to assess knee OA severity using radiography-based semi-quantitative scoring systems, like KL gradings, which are based on the study of anatomical features such as variations in joint space width or osteophytes formation \cite{KneeOA3} \cite{KneeOA2} \cite{KneeOA4}. Shamir et al. \cite{KneeOA4} proposed WND-CHARM: a multi purpose medical image classifier to automatically assess knee OA severity in radiographs using a set of features based on polynomial decompositions, contrast, pixel statistics, textures, and features from image transforms. Recently, Yoo et al. \cite{KneeOA6} proposed a self-assessment scoring system associating risk factors and radiographic knee OA features using multivariable logistic regression models, additionally using an Artificial Neural Network (ANN) to improve the overall scoring performance. Shamir et. al. \cite{KneeOA2} proposed template matching to automatically detect knee joints from X-ray images. This method is slow to compute for large datasets and gives poor detection performance. Antony et al. \cite{antony2} introduced an SVM-based approach for automatically detecting the knee joints. Later, Antony et al. \cite{antony2017automatic} proposed an FCN-based approach to improve the localization of the knee joints. Although more accurate, the aspect ratio chosen for the extracted knee joints affects the overall quantification. 

Recently, the emergence of deep learning has enabled the development of new intelligent diagnostics based on computer vision. CNNs outperform many state-of-the-art methods based on hand-crafted features in tasks such as image classification \cite{krizhevsky2012imagenet}, retrieval \cite{babenko2014neural} and object detection \cite{lawrence1997face} \cite{wei2011computer}. Antony et al. \cite{antony2} showed that the off-the-shelf CNNs such as the VGG 16-layer network \cite{VGG16}, the VGG-M-128 network \cite{chatfield2014return}, and the BVLC reference CaffeNet \cite{jia2014caffe} \cite{karayev2013recognizing} pre-trained on ImageNet LSVRC dataset \cite{russakovsky2015imagenet} can be fine-tuned for classifying knee OA images through transfer learning. They argued that it is appropriate to assess knee OA severity using continuous metrics like mean-squared error together with binary or multi-class classification losses, showing that predicting the continuous grades through regression reduces the error and improves overall quantification. They proposed a novel pipeline \cite{antony2017automatic} to automatically quantify knee OA severity using a FCN for localization  and a CNN jointly trained for classification and regression. The work consolidates the state-of-the-art baseline for the application of CNNs in the field, opening a range of research lines for further improvements. Tiulpin et al. \cite{tiulpin2018automatic} presented a new computer-aided diagnosis method based on using deep Siamese CNNs, which are originally designed to learn a similarity metric between pairs of images. However, rather than comparing image pairs, the authors extend this idea to similarity in knee x-ray images (with 2 symmetric knee joints). Splitting the images at the central position and feeding both knee joints into a separate CNN branch allows the network to learn identical weights for both branches. They outperform the previous approaches by achieving an average multi-class testing accuracy score of 66.71 \% on the entire OAI dataset, despite also needing  a localization step to focus the network branches on the knee joint areas.

This work mainly focuses on designing an end-to-end architecture with attention mechanisms. There are similar methods reported in the literature. Xiao et al. \cite{xiao2015application} propose a pipeline to apply visual attention to deep neural networks by integrating and combining attention models to train domain-specific nets. In another approach,  Liu et al. \cite{liu2016fully} introduce a reinforcement learning framework based on fully convolutional attention networks (FCAN) to optimally select local discriminative regions adaptive to different fine-grained domains. The proposed weakly-supervised reinforcement method combined with a fully-convolutional architecture achieves fast convergence without requiring expensive annotation. Recently, Jetley et al. \cite{jetley2018learn} introduce an end-to-end-trainable attention module for CNN architectures built for image classification. The module takes as input the 2D feature vector maps, which forms the intermediate representations of the input image at different stages in the CNN pipeline, and outputs a matrix of scores for each map. They redesign standard architectures to classify the input image using only a weighted combination of local features, forcing the network to learn relevant attention patterns.

\section{Method}
This section describes the proposed methods, detailing the design, implementation, and training of the attention modules. Several strategies are investigated to integrate the attention mechanism into standard CNN architectures, proposing experimental approaches to classify the knee images.

\subsection{Trainable Attention Module for CNNs}
\begin{figure}[ht]
\floatconts
  {fig:example}
  {\caption{Attention module scheme.}}
  {\includegraphics[width=0.8\textwidth, height=0.32\textwidth]{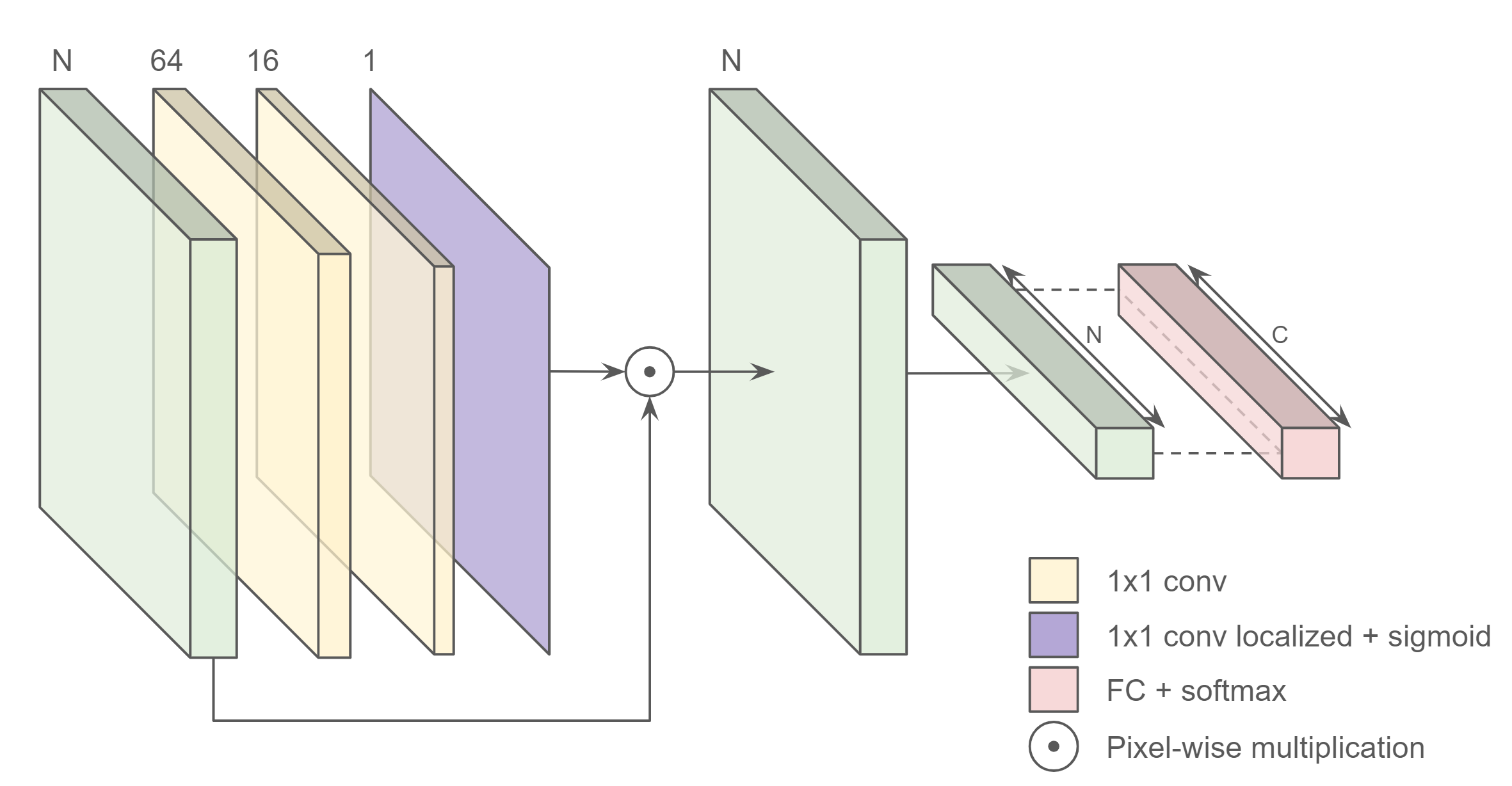}}
  \label{fig:att}
\end{figure}

The selected attention module is inspired by the work of Kevin Mader \cite{kevinmader}, in which a trainable attention mechanism is designed for a pretrained VGG-16 network to predict bone age from hand X-Ray images.
Figure \ref{fig:att} illustrates this idea. 

Given an input volume $D^{l}$ from a convolutional layer $l$ with $N$ feature maps, several $1\times 1$ convolutional layers are stacked to extract spatial features. The output is then passed to a $1\times 1$ locally connected layer \cite{chen2015locally} (convolution with unshared weights) with sigmoidal activation to give an attention mask $A^{l}$. 
The original feature maps are element-wise multiplied by the attention mask, generating a new convolutional volume $\tilde{D}^{l}$ accentuating informative areas.
A spatial dimensionality reduction is performed by applying global average pooling (GAP) on the masked volume, generating a $N$-dimensional feature vector $F^{l}$, which is then normalized by the average value of the attention mask. Additionally, a \textit{softmax} layer can be applied to yield a $C$-dimensional vector with the output class probabilities $\lbrace p_{1}, p_{2}, \ ..., \ p_{C} \rbrace$.

\subsection{Module Integration to CNN Pipeline}
Standard CNN architectures typically stack several convolutional layers with occasional pooling operations that reduces the spatial dimension and increase the receptive field. Therefore, the degree of abstraction of the attention modules is closely related to their location in the CNN, focusing on more global details as depth increases. We define the concept of an attention branch as the location of an attention module in a specific convolutional block, applying a \textit{softmax} operation at the top to produce independent class probabilities based on the KL scores. Each attention branch will be seen as a new model by itself that could be trained end-to-end. Figure \ref{fig:vgg-example} shows a sample architecture integrating the attention modules to the VGG-16 pipeline. Fixing an input size of $320\times224$ pixels, as fixed in Section \ref{sec:data}, we build the branches $att\{i\}_{i=0..2}$ taking as input volumes the feature maps belonging to the pooling layers after the convolutional blocks $3$, $4$, and $5$. Following the methodology in Section \ref{sec:combination}, a combinational module is applied to fuse the local features from all the branches into a global feature vector and then generate the KL grades probability distribution by applying a \textit{softmax} layer at the top.



\begin{figure}[ht]
\floatconts
  {fig:vgg-example}
  {\caption{Sample architecture integrating the attention modules in the VGG-16 pipeline.}}
  {\includegraphics[width=0.98\textwidth, height=0.45\textwidth]{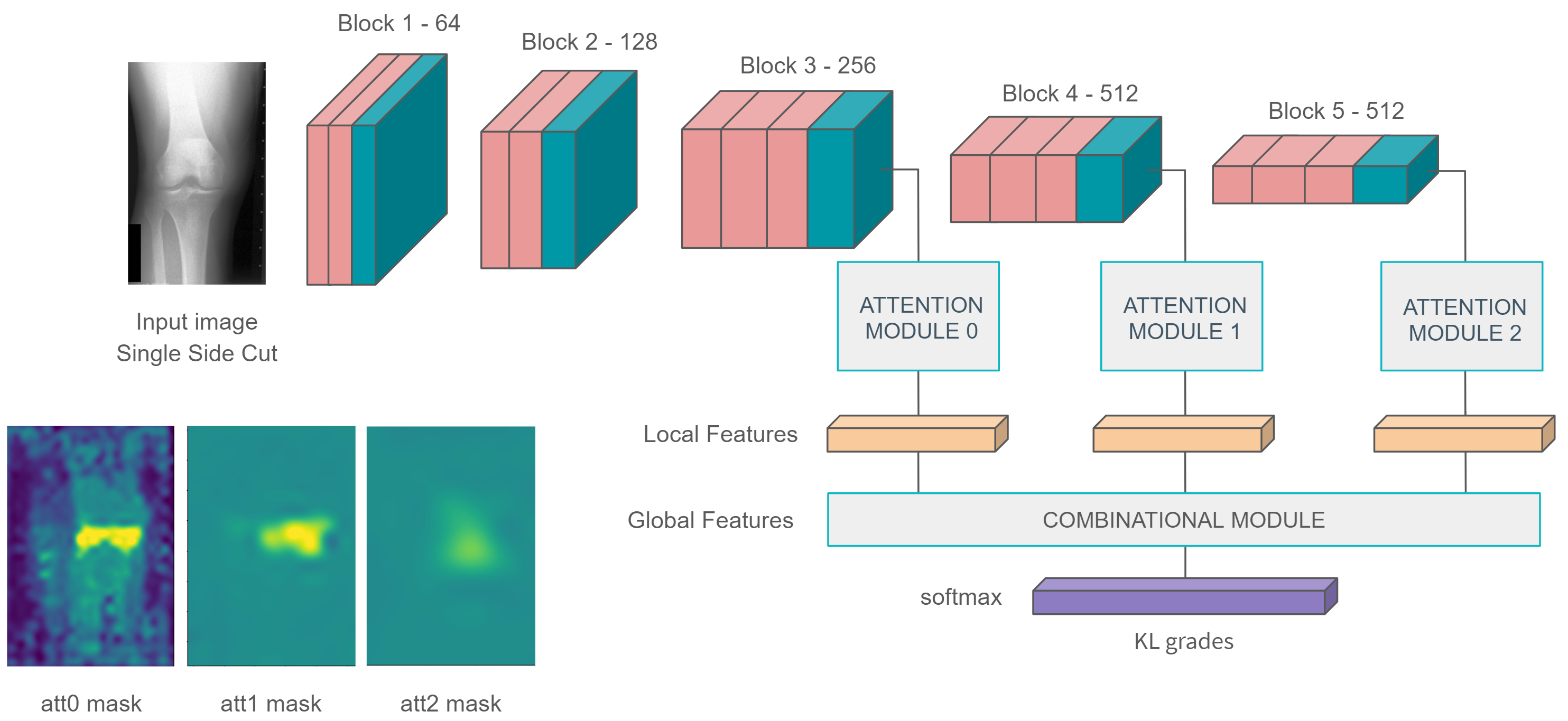}}
\end{figure}

\subsection{Combining Multiple Attention Branches}
\label{sec:combination}
Several strategies are investigated to merge features from multiple branches with the aim to combine attention patterns at different resolutions.  Our first strategy is performing early fusion of features from different branches. Each attention module generates a $N$-dimensional feature vector $F^{l}$ with the average values of the $N$ masked feature maps conforming the input convolutional volume. A channel-wise concatenation is applied to fuse all the branches, generating a new vector $F_{A} = [F^{l_{1}} \ F^{l_{2}} \ ... \ F^{l_{B}}]$ with $F_{A} \in R^{1x(\sum{N_{b})} }$, being $B$ the total of attention branches and $N_{b}$ the dimension of $F^{l_{b}}$. In addition, a fully connected layer is added at the top to perform early fusion of the concatenated features, while a \textit{softmax} operation is applied to generate the $C$-dimensional output class probabilities. As shown above, the complexity of the attention modules correlates with their location in the CNN pipeline, which biases the convergence behavior. This can be critical for a combined model that attempts to train modules with different convergence rates at the same time: deeper branches quickly overfit while waiting for the convergence of the slower ones. In contrast, by reducing the overall learning time, the shallower branches with more  complex modules may decrease their performance due to under training.

Our next strategy is to simplify the multiple branch learning process. We propose the use of multi-loss training, which aims to improve the learning efficiency and prediction performance by learning multiple objectives from a shared representation. Each attention branch makes separate predictions via a \textit{softmax} to to generate their class probabilities and we linearly combine the individual categorical cross-entropies $\mathcal{L}_{b}$ into a global loss: $\mathcal{L} = \sum_{b}{w}_{b}\mathcal{L}_{b}$, with $\mathcal{L}_{b} = -\sum_{c}y_{c}^{(b)}\log p_{c}^{(b)}\label{eq:2}$. This allows to control the rate of convergence by weighting the contribution of each branch, assigning low weights for those branches with faster convergence to reduce their influence at the initial stages of training and attenuate updates in shallower attention modules. There are previous approaches that propose the use of multi-loss training to address different machine learning tasks such as dense prediction \cite{kokkinos2017ubernet}, scene understanding \cite{eigen2015predicting}, natural language processing \cite{collobert2008unified} or speech recognition \cite{huang2013cross}. However, the model performance is extremely sensitive to the weight selection ${w}_{b}$, that needs an expensive and time-consuming hyper-parametrization process.

Several multi-branch combinations were tested by applying multidimensional cross-validation to find the optimum branch locations and multi-loss weights. We used a 2D grid search, validating the $att0$ and $att1$ loss weights between a range of $0.5$ to $1$ with a step size of $0.1$, and using the validation loss as monitor. The best performance was achieved with $att0$, $att1$ weights ${w}_{0}=1$, ${w}_{1}=0.8$, slightly reducing the contribution of the deeper attention modules and decreasing their overfitting tendency while the shallower branches are still learning.

\begin{figure}[h]
\floatconts
  {fig:merge}
  {\caption{Comparison between merging solutions in the VGG-16 pipeline, visualizing the generated masks in the attention branches $att0$ and $att1$ and observing a large improvement in the shallower modules using multi-loss.}}
  {\includegraphics[width=0.5\textwidth]{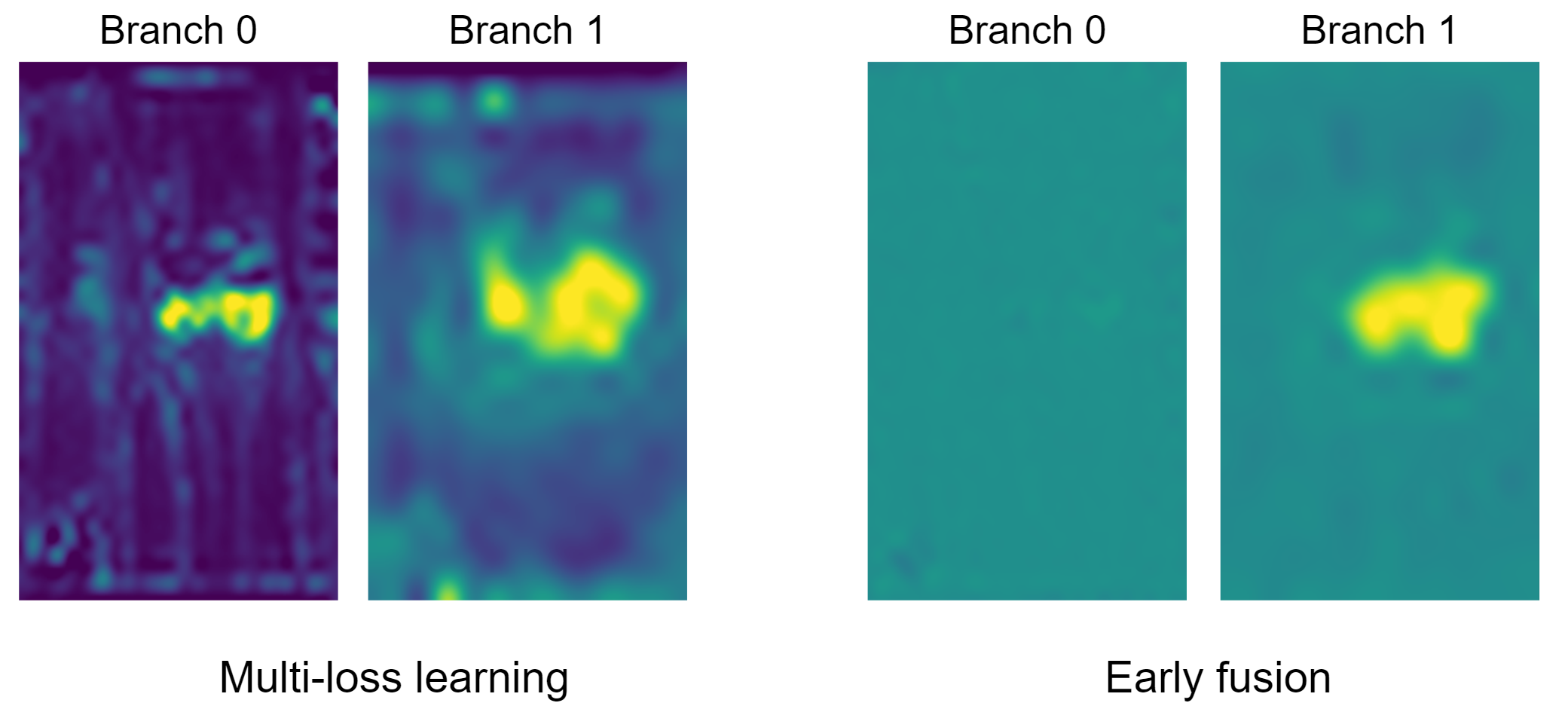}}
\end{figure}

\subsection{Public Knee OA Datasets}
\label{sec:data}
The data used for this work are bilateral PA fixed flexion knee X-ray images. The datasets are from the Osteoarthritis Initiative (OAI) and Multicenter Osteoarthritis Study (MOST) in UCSF, being standard public datasets widely used in knee OA studies. The baseline cohort of the OAI dataset contains MRI and X-ray images of 4,476 participants. From this entire cohort, we selected 4,446 X-ray images based on the availability of KL grades for both knees as per the assessments by Boston University X-ray reading center (BU). The MOST dataset includes lateral knee radiograph assessments of 3,026 participants. From this, 2,920 radiographs are selected based on the availability of KL grades for both knees as per baseline assessments. As a pre-processing step, all the X-ray images are manually split in the middle, generating two vertical sections from the left and right sides, resizing them to a fixed mean size of $320 \times 224$ pixels by keeping the average aspect ratio. Histogram equalization is performed for intensity level normalization, and eventually data augmentation is applied by performing horizontal right-left flips to generate more training data. The training, validation, and test sets were split based on the KL grades distribution. A 70-30  train-test split was used and $10\%$ of the training data was kept for validation.

\subsection{Training}
All models were trained from scratch using categorical cross-entropy with the ground truth KL grades. Regarding multi-branch training, all target data were duplicated for each attention branch. We used Adam \cite{kingma2014adam} with a batch size of 64, ${\beta}_{1}=0.9$, ${\beta}_{2}=0.999$, an initial learning rate of $10^{-5}$ scaled by $0.1$ every $2$ epochs without improvement in validation loss, and early stopping after $3$ epochs without improvement.

\section{Results}
Although the attention mechanism can be integrated to any CNN pipeline, not all the architectures are well-suited for assessing knee OA severity. We explored several architectures in the literature including state-of-the-art models from previous works of Antony et al. \cite{antony2017automatic} and more complex architectures such as ResNet-50 \cite{Resnet50}, and we analyzed their performance. We found that the same level of abstraction in $att1$ and $att2$ for Antony et. al models can be achieved in shallower branches $att0$, $att1$ for deeper architectures, implying the best branch location depends on model complexity. After testing different branch combinations, the best performing locations presented in Table \ref{tab:evaluation}, and detailed in the Tables \ref{tab:antony-clf}, \ref{tab:antony-ext}, \ref{tab:resnet50}, specifying the output resolution of their convolutional blocks and the location of the attention branches. From the evaluation of multi-loss models, since every single attention branch produces an independent prediction, the top performing one is used at test time. A more sophisticated ensemble approach was considered but not included in this paper. This approach involves averaging the pre-activation outputs (i.e. values before the softmax) of each of the model branches and then passing the result through a softmax to enforce a valid probability distribution over the classes. This idea is often effective in test time data augmentation and ensemble methods and may improve performance over the single best model referred here. As Table \ref{tab:evaluation} shows, the VGG-16 attention branch $att0$ with multi-loss learning achieved the best overall classification accuracy ($64.3\%$).

\begin{table}[htbp]
\floatconts{tab:evaluation}
  {\caption{Evaluation overview for different CNN pipelines.}}%
  {\renewcommand{\arraystretch}{1.3}\begin{tabular}{c|c|c|c|c}
                    \bfseries Models& \bfseries Antony Clsf.     & \bfseries Antony Extended       & \bfseries ResNet-50        & \bfseries VGG-16          \\ \hline
$att0$                & $41.76\%$        & $40\%$            & $59.3\%$          & $62.2\%$         \\
$att1$              & $41.9\%$         & $53.26\%$         & $58.67\%$         & $61.7\%$         \\
$att2$              & $44.53\%$        & $53.83\%$         & $56.47\%$         & $58.8\%$         \\ \hline
\multirow{1}{*}{Early fusion} & $52.5\%$  & $56.17\%$ & $59\%$  & $63\%$  \\ \hline
\multirow{2}{*}{Multi-Loss} & $att1$: $43.9\%$ & $att1$: $55.61\%$ & $att0$: \textbf{60\%}    & $att0$: \textbf{64.3\%} \\
                            & $att2$: $46.6\%$ & $att2$: $55.68\%$ & $att1$: $56.88\%$ & $att1$: $63.2\%$ \\ \hline
\end{tabular}}
\end{table}

We also compared the attention mechanism with related knee OA classification-based solutions in the literature. First, we retrained the Antony et. al models with the same training data from the previous experiments, applying the FCN introduced by the authors to address the knee joints extraction \cite{antony2017automatic}. The results (Table \ref{tab:comparative}) show that the attention-based models with end-to-end architectures clearly outperform the state-of-the-art frameworks. We further compared our results to human level accuracy using the radiologic reliability readings from the OAI \cite{klara2016reliability}. Although the data used to compute the reliability grading does not match our test set, we followed the methodology of previous works in the literature \cite{tiulpin2018automatic}, with the aim to dispose of a panoramic view of the current gold standard for diagnosing OA involving human performance. Cohen's \textit{kappa} coefficient \cite{cohen1960coefficient} was used to evaluate the agreement between non-clinician readers and experienced radiologists by classifying $N$ items with $C$ mutually exclusive categories. Considering the following grading: $\kappa <0.20$ slight agreement, $0.21$-$0.40$ fair, $0.41$-$0.60$ moderate, $0.61$-$0.80$ substantial, and $0.81$-$1.0$ almost perfect agreement, their inter-reader reliability for the KL scores was moderate to substantial, with $\kappa$ values between $0.5$ and $0.8$. In the case of automatic assessments, by considering a CNN model as a non-clinician X-Ray reader, we can apply the $\kappa$ coefficient to evaluate the inter-reader reliability between its predictions and the corresponding ground truth annotations, provided by experienced radiologists. As Table \ref{tab:comparative} shows, the $att0$ branch of the VGG-16 trained by multi-loss together with the $att1$ branch ($w_{1}=0.8$), improves the reliability of related works with a substantial $\kappa = 0.63$ agreement, reaching the margins of human accuracy.

\begin{table}[htbp]
\floatconts{tab:comparative}%
  {\caption{Comparative with related frameworks}}%
  {\adjustbox{width=\textwidth}{\renewcommand{\arraystretch}{1.3}\begin{tabular}{r|c|c|c|c}
                          & \bfseries Test Acc. & \bfseries Test Loss & \bfseries Kappa  & \bfseries \# parameters                                      \\\hline
Antony Clsf              & $59\%$    & $1.13$    & $0.42$ & $\sim 7.51$ M (FCN: $\sim212.7$ K)  \\
Antony Joint Clsf \& Reg & $62.29\%$ & $0.89$    & $0.48$ & $\sim 5.77$ M (FCN: $\sim212.7$ K) \\
VGG-16: Multi-loss ($att0$)        & \textbf{64.3\%}  & $0.817$   & \textbf{0.63} & $\sim7.7$ M    \\\hline                                      
\end{tabular}}}
\end{table}

\section{Conclusions}

This work proposed a novel end-to-end architecture that incorporates trainable attention modules acting as unsupervised fine-grained ROI detectors. The proposed attention modules can be applied at different levels and scales across the CNN pipeline, helping the network to learn relevant attention patterns over the most informative parts of the image at different resolutions. The results obtained for the public knee OA datasets OAI and MOST were satisfactory despite having a considerable scope for further improvement.

The proposed attention mechanism can be easily integrated to any convolutional neural network architecture, being adaptable to any input convolutional volume. However, after exploring different off-the-shelf base models for classification with different complexities, we observed that the best performance is achieved in those models with a balanced ratio between the complexity of the overall architecture and the depth of the convolutional volumes, avoiding overfitting while getting abstraction in the local features used to train the attention modules. On the other hand, we propose the use of multi-loss training to manage the training of multiple attention branches with different velocities of convergence at the same time, boosting the overall performance by fusing attention features with different levels of abstraction. The best performance was achieved by slightly reducing the contribution of the deepest attention branches, improving then the precision of the shallower attention masks and reaching the effectiveness of related approaches with a test accuracy of $64.3 \%$ and Kappa agreement of $0.63$. Although our method does not surpass the state-of-the-art and could be interpreted as challenging to implement, the overall aim was to reduce the training complexity using an end-to-end architecture. As mentioned in Section 4, without an end-to-end design, the models require a localization step to focus the classifier to the knee joint regions of interest. For instance, previous work of Antony et. al. \cite{antony2017automatic} needed a manual annotation process for training a FCN to automatically segment the input knee joints. Our approach, in contrast, requires no such annotation of knee joint locations in the training data. Finally, we observed that localizing the knee joints in an unsupervised way can reduce performance by adding noise in the attention masks and thus into the overall process. A more robust attention module can improve the results and have a bigger impact in the future. As future work, it may be interesting to design better base networks for the attention mechanism and then to test new fine-grained methods in the state-of-art, with the aim to improve the performance of the attention modules towards reducing their dependence on the complexity of the base model. 

\midlacknowledgments{This research was supported by contract SGR1421 by the Catalan AGAUR office. The work has been developed in the framework of project TEC2016-75976-R, funded by the Spanish Ministerio de Economia y Competitividad and the European Regional Development Fund (ERDF). The authors also thank NVIDIA for generous hardware donations.~

This publication has emanated from research conducted with the financial support of Science Foundation Ireland (SFI) under grant numbers SFI/12/RC/2289 and 15/SIRG/3283.~

The OAI is a public-private partnership comprised of five contracts (N01-AR-2-2258; N01-AR-2-2259; N01-AR-2- 2260; N01-AR-2-2261; N01-AR-2-2262) funded by the National Institutes of Health, a branch of the Department of Health and Human Services, and conducted by the OAI Study Investigators. Private funding partners include Merck Research Laboratories; Novartis Pharmaceuticals Corporation, GlaxoSmithKline; and Pfizer, Inc. Private sector funding for the OAI is managed by the Foundation for the National Institutes of Health. MOST is comprised of four cooperative grants (Felson -- AG18820; Torner -- AG18832; Lewis -- AG18947; and Nevitt -- AG19069) funded by the National Institutes of Health, a branch of the Department of Health and Human Services, and conducted by MOST study investigators. This manuscript was prepared using MOST data and does not necessarily reflect the opinions or views of MOST investigators.}

\bibliography{Gorriz19}

\begin{thebibliography}{33}
\providecommand{\natexlab}[1]{#1}
\providecommand{\url}[1]{\texttt{#1}}
\expandafter\ifx\csname urlstyle\endcsname\relax
  \providecommand{\doi}[1]{doi: #1}\else
  \providecommand{\doi}{doi: \begingroup \urlstyle{rm}\Url}\fi

\bibitem[Antony et~al.(2016)Antony, McGuinness, O'Connor, and Moran]{antony2}
Joseph Antony, Kevin McGuinness, Noel~E O'Connor, and Kieran Moran.
\newblock Quantifying radiographic knee osteoarthritis severity using deep
  convolutional neural networks.
\newblock In \emph{Pattern Recognition (ICPR), 2016 23rd International
  Conference on}, pages 1195--1200. IEEE, 2016.

\bibitem[Antony et~al.(2017)Antony, McGuinness, Moran, and
  O’Connor]{antony2017automatic}
Joseph Antony, Kevin McGuinness, Kieran Moran, and Noel~E O’Connor.
\newblock Automatic detection of knee joints and quantification of knee
  osteoarthritis severity using convolutional neural networks.
\newblock In \emph{International Conference on Machine Learning and Data Mining
  in Pattern Recognition}, pages 376--390. Springer, 2017.

\bibitem[Babenko et~al.(2014)Babenko, Slesarev, Chigorin, and
  Lempitsky]{babenko2014neural}
Artem Babenko, Anton Slesarev, Alexandr Chigorin, and Victor Lempitsky.
\newblock Neural codes for image retrieval.
\newblock In \emph{European conference on computer vision}, pages 584--599.
  Springer, 2014.

\bibitem[Chatfield et~al.(2014)Chatfield, Simonyan, Vedaldi, and
  Zisserman]{chatfield2014return}
Ken Chatfield, Karen Simonyan, Andrea Vedaldi, and Andrew Zisserman.
\newblock Return of the devil in the details: Delving deep into convolutional
  nets.
\newblock \emph{arXiv preprint arXiv:1405.3531}, 2014.

\bibitem[Chen et~al.(2015)Chen, Lopez-Moreno, Sainath, Visontai, Alvarez, and
  Parada]{chen2015locally}
Yu-hsin Chen, Ignacio Lopez-Moreno, Tara~N Sainath, Mirk{\'o} Visontai, Raziel
  Alvarez, and Carolina Parada.
\newblock Locally-connected and convolutional neural networks for small
  footprint speaker recognition.
\newblock In \emph{Sixteenth Annual Conference of the International Speech
  Communication Association}, 2015.

\bibitem[Cohen(1960)]{cohen1960coefficient}
Jacob Cohen.
\newblock A coefficient of agreement for nominal scales.
\newblock \emph{Educational and psychological measurement}, 20\penalty0
  (1):\penalty0 37--46, 1960.

\bibitem[Collobert and Weston(2008)]{collobert2008unified}
Ronan Collobert and Jason Weston.
\newblock A unified architecture for natural language processing: Deep neural
  networks with multitask learning.
\newblock In \emph{Proceedings of the 25th international conference on Machine
  learning}, pages 160--167. ACM, 2008.

\bibitem[Eigen and Fergus(2015)]{eigen2015predicting}
David Eigen and Rob Fergus.
\newblock Predicting depth, surface normals and semantic labels with a common
  multi-scale convolutional architecture.
\newblock In \emph{Proceedings of the IEEE International Conference on Computer
  Vision}, pages 2650--2658, 2015.

\bibitem[He et~al.(2016)He, Zhang, Ren, and Sun]{Resnet50}
Kaiming He, Xiangyu Zhang, Shaoqing Ren, and Jian Sun.
\newblock Deep residual learning for image recognition.
\newblock In \emph{Proceedings of the IEEE conference on computer vision and
  pattern recognition}, pages 770--778, 2016.

\bibitem[Heidari(2011)]{heidari2011knee}
Behzad Heidari.
\newblock Knee osteoarthritis prevalence, risk factors, pathogenesis and
  features: Part i.
\newblock \emph{Caspian journal of internal medicine}, 2\penalty0 (2):\penalty0
  205, 2011.

\bibitem[Huang et~al.(2013)Huang, Li, Yu, Deng, and Gong]{huang2013cross}
Jui-Ting Huang, Jinyu Li, Dong Yu, Li~Deng, and Yifan Gong.
\newblock Cross-language knowledge transfer using multilingual deep neural
  network with shared hidden layers.
\newblock In \emph{Acoustics, Speech and Signal Processing (ICASSP), 2013 IEEE
  International Conference on}, pages 7304--7308. IEEE, 2013.

\bibitem[Jetley et~al.(2018)Jetley, Lord, Lee, and Torr]{jetley2018learn}
Saumya Jetley, Nicholas~A Lord, Namhoon Lee, and Philip~HS Torr.
\newblock Learn to pay attention.
\newblock \emph{arXiv preprint arXiv:1804.02391}, 2018.

\bibitem[Jia et~al.(2014)Jia, Shelhamer, Donahue, Karayev, Long, Girshick,
  Guadarrama, and Darrell]{jia2014caffe}
Yangqing Jia, Evan Shelhamer, Jeff Donahue, Sergey Karayev, Jonathan Long, Ross
  Girshick, Sergio Guadarrama, and Trevor Darrell.
\newblock Caffe: Convolutional architecture for fast feature embedding.
\newblock In \emph{Proceedings of the 22nd ACM international conference on
  Multimedia}, pages 675--678. ACM, 2014.

\bibitem[Karayev et~al.(2013)Karayev, Trentacoste, Han, Agarwala, Darrell,
  Hertzmann, and Winnemoeller]{karayev2013recognizing}
Sergey Karayev, Matthew Trentacoste, Helen Han, Aseem Agarwala, Trevor Darrell,
  Aaron Hertzmann, and Holger Winnemoeller.
\newblock Recognizing image style.
\newblock \emph{arXiv preprint arXiv:1311.3715}, 2013.

\bibitem[Kellgren and Lawrence(1957)]{KL}
JH~Kellgren and JS~Lawrence.
\newblock Radiological assessment of osteo-arthrosis.
\newblock \emph{Annals of the rheumatic diseases}, 16\penalty0 (4):\penalty0
  494, 1957.

\bibitem[Kingma and Ba(2014)]{kingma2014adam}
Diederik~P Kingma and Jimmy Ba.
\newblock Adam: A method for stochastic optimization.
\newblock \emph{arXiv preprint arXiv:1412.6980}, 2014.

\bibitem[Klara et~al.(2016)Klara, Collins, Gurary, Elman, Stenquist, Losina,
  and Katz]{klara2016reliability}
Kristina Klara, Jamie~E Collins, Ellen Gurary, Scott~A Elman, Derek~S
  Stenquist, Elena Losina, and Jeffrey~N Katz.
\newblock Reliability and accuracy of cross-sectional radiographic assessment
  of severe knee osteoarthritis: role of training and experience.
\newblock \emph{The Journal of rheumatology}, pages jrheum--151300, 2016.

\bibitem[Kokkinos(2017)]{kokkinos2017ubernet}
Iasonas Kokkinos.
\newblock Ubernet: Training a universal convolutional neural network for low-,
  mid-, and high-level vision using diverse datasets and limited memory.
\newblock In \emph{CVPR}, volume~2, page~8, 2017.

\bibitem[Krizhevsky et~al.(2012)Krizhevsky, Sutskever, and
  Hinton]{krizhevsky2012imagenet}
Alex Krizhevsky, Ilya Sutskever, and Geoffrey~E Hinton.
\newblock Imagenet classification with deep convolutional neural networks.
\newblock In \emph{Advances in neural information processing systems}, pages
  1097--1105, 2012.

\bibitem[Lawrence et~al.(1997)Lawrence, Giles, Tsoi, and
  Back]{lawrence1997face}
Steve Lawrence, C~Lee Giles, Ah~Chung Tsoi, and Andrew~D Back.
\newblock Face recognition: A convolutional neural-network approach.
\newblock \emph{IEEE transactions on neural networks}, 8\penalty0 (1):\penalty0
  98--113, 1997.

\bibitem[Li et~al.(2013)Li, Wei, Zhou, and Wei]{KneeOA5}
YongPing Li, XiaoChun Wei, JingMing Zhou, and Lei Wei.
\newblock The age-related changes in cartilage and osteoarthritis.
\newblock \emph{BioMed research international}, 2013, 2013.

\bibitem[Liu et~al.(2016)Liu, Xia, Wang, Yang, Zhou, and Lin]{liu2016fully}
Xiao Liu, Tian Xia, Jiang Wang, Yi~Yang, Feng Zhou, and Yuanqing Lin.
\newblock Fully convolutional attention networks for fine-grained recognition.
\newblock \emph{arXiv preprint arXiv:1603.06765}, 2016.

\bibitem[Mader(2018)]{kevinmader}
Kevin Mader.
\newblock Attention on pretrained-vgg16 for bone age.
\newblock
  \url{https://www.kaggle.com/kmader/attention-on-pretrained-vgg16-for-bone-age},
  2018.

\bibitem[Oka et~al.(2008)Oka, Muraki, Akune, Mabuchi, Suzuki, Yoshida,
  Yamamoto, Nakamura, Yoshimura, and Kawaguchi]{KneeOA1}
H~Oka, S~Muraki, T~Akune, A~Mabuchi, T~Suzuki, H~Yoshida, S~Yamamoto,
  K~Nakamura, N~Yoshimura, and H~Kawaguchi.
\newblock Fully automatic quantification of knee osteoarthritis severity on
  plain radiographs.
\newblock \emph{Osteoarthritis and Cartilage}, 16\penalty0 (11):\penalty0
  1300--1306, 2008.

\bibitem[Russakovsky et~al.(2015)Russakovsky, Deng, Su, Krause, Satheesh, Ma,
  Huang, Karpathy, Khosla, Bernstein, et~al.]{russakovsky2015imagenet}
Olga Russakovsky, Jia Deng, Hao Su, Jonathan Krause, Sanjeev Satheesh, Sean Ma,
  Zhiheng Huang, Andrej Karpathy, Aditya Khosla, Michael Bernstein, et~al.
\newblock Imagenet large scale visual recognition challenge.
\newblock \emph{International Journal of Computer Vision}, 115\penalty0
  (3):\penalty0 211--252, 2015.

\bibitem[Shamir et~al.(2008)Shamir, Orlov, Eckley, Macura, Johnston, and
  Goldberg]{KneeOA4}
Lior Shamir, Nikita Orlov, D~Mark Eckley, Tomasz Macura, Josiah Johnston, and
  Ilya~G Goldberg.
\newblock Wndchrm--an open source utility for biological image analysis.
\newblock \emph{Source code for biology and medicine}, 3\penalty0 (1):\penalty0
  13, 2008.

\bibitem[Shamir et~al.(2009{\natexlab{a}})Shamir, Ling, Scott, Hochberg,
  Ferrucci, and Goldberg]{KneeOA2}
Lior Shamir, Shari~M Ling, William Scott, Marc Hochberg, Luigi Ferrucci, and
  Ilya~G Goldberg.
\newblock Early detection of radiographic knee osteoarthritis using
  computer-aided analysis.
\newblock \emph{Osteoarthritis and Cartilage}, 17\penalty0 (10):\penalty0
  1307--1312, 2009{\natexlab{a}}.

\bibitem[Shamir et~al.(2009{\natexlab{b}})Shamir, Ling, Scott~Jr, Bos, Orlov,
  Macura, Eckley, Ferrucci, and Goldberg]{KneeOA3}
Lior Shamir, Shari~M Ling, William~W Scott~Jr, Angelo Bos, Nikita Orlov,
  Tomasz~J Macura, D~Mark Eckley, Luigi Ferrucci, and Ilya~G Goldberg.
\newblock Knee x-ray image analysis method for automated detection of
  osteoarthritis.
\newblock \emph{IEEE Transactions on Biomedical Engineering}, 56\penalty0
  (2):\penalty0 407--415, 2009{\natexlab{b}}.

\bibitem[Simonyan and Zisserman(2014)]{VGG16}
Karen Simonyan and Andrew Zisserman.
\newblock Very deep convolutional networks for large-scale image recognition.
\newblock \emph{arXiv preprint arXiv:1409.1556}, 2014.

\bibitem[Tiulpin et~al.(2018)Tiulpin, Thevenot, Rahtu, Lehenkari, and
  Saarakkala]{tiulpin2018automatic}
Aleksei Tiulpin, J{\'e}r{\^o}me Thevenot, Esa Rahtu, Petri Lehenkari, and Simo
  Saarakkala.
\newblock Automatic knee osteoarthritis diagnosis from plain radiographs: a
  deep learning-based approach.
\newblock \emph{Scientific reports}, 8\penalty0 (1):\penalty0 1727, 2018.

\bibitem[Wei et~al.(2011)Wei, Chan, Zhou, Wu, Sahiner, Hadjiiski, Roubidoux,
  and Helvie]{wei2011computer}
Jun Wei, Heang-Ping Chan, Chuan Zhou, Yi-Ta Wu, Berkman Sahiner, Lubomir~M
  Hadjiiski, Marilyn~A Roubidoux, and Mark~A Helvie.
\newblock Computer-aided detection of breast masses: Four-view strategy for
  screening mammography.
\newblock \emph{Medical physics}, 38\penalty0 (4):\penalty0 1867--1876, 2011.

\bibitem[Xiao et~al.(2015)Xiao, Xu, Yang, Zhang, Peng, and
  Zhang]{xiao2015application}
Tianjun Xiao, Yichong Xu, Kuiyuan Yang, Jiaxing Zhang, Yuxin Peng, and Zheng
  Zhang.
\newblock The application of two-level attention models in deep convolutional
  neural network for fine-grained image classification.
\newblock In \emph{Proceedings of the IEEE Conference on Computer Vision and
  Pattern Recognition}, pages 842--850, 2015.

\bibitem[Yoo et~al.(2016)Yoo, Kim, Choi, and Park]{KneeOA6}
Tae~Keun Yoo, Deok~Won Kim, Soo~Beom Choi, and Jee~Soo Park.
\newblock Simple scoring system and artificial neural network for knee
  osteoarthritis risk prediction: a cross-sectional study.
\newblock \emph{PloS one}, 11\penalty0 (2):\penalty0 e0148724, 2016.

\end{thebibliography}

\newpage
\appendix
\section{CNN Architectures, Learning Curves and Visualizations}

\begin{table}[ht]
\floatconts
  {tab:antony-clf}%
  {\caption{Antony et al. base architecture for classification}}%
  {\renewcommand{\arraystretch}{1.3}\begin{tabular}{r|c|c|c|c}
\bfseries Layer    & \bfseries Kernels & \bfseries Kernel Size & \bfseries Strides & \bfseries Output shape \\ \hline
conv1    & $32$      & $11 \times 11$       & $2$       & $100 \times 150 \times 32$   \\
pool1 & -       & $3 \times 3$         & $2$       & $49 \times 74 \times 32$     \\\hline
conv2    & $64$      & $5 \times 5$         & $1$       & $49 \times 74 \times 64$     \\
$(att0)$ pool2 & -       & $3 \times 3$         & $2$       & $24 \times 36 \times 64$     \\\hline
conv3    & $96$      & $3 \times 3$         & $1$       & $24 \times 36 \times 96$     \\
$(att1)$ pool3  & -       & $3 \times 3$         & $2$       & $11 \times 17 \times 96$     \\\hline
conv4    & $128$     & $3 \times 3$         & $1$       & $11 \times 17 \times 128$    \\
$(att2)$ pool4  & -       & $3 \times 3$         & $2$       & $5 \times 8 \times 128$      \\\hline
\end{tabular}}
\end{table}

\begin{table}[ht]
\floatconts
  {tab:antony-ext}%
  {\caption{Antony et al. extended base architecture for classification and regression}}%
  {\renewcommand{\arraystretch}{1.3}\begin{tabular}{r|c|c|c|c}
\bfseries Layer                        & \bfseries Kernels & \bfseries Kernel Size & \bfseries Strides & \bfseries Output shape \\ \hline
conv1                        & $32$      & $11 \times  11$       & $2$       & $100 \times  150 \times  32$   \\
pool1                     & -       & $3 \times  3$         & $2$       & $49 \times  74 \times  32$     \\\hline
conv2-1                      & $64$     & $3 \times  3$          & $1$       & $49 \times  74 \times  64$     \\
conv2-2                      & $64$      & $3 \times  3$         & $1$      & $49 \times  74 \times  64$     \\
$(att0)$ pool2    & -       & $3 \times  3$         & $2$      & $24 \times  36 \times  64$     \\\hline
conv3-1                      & $96$      & $3 \times  3$         & $1$       & $24 \times  36 \times  96$     \\
conv3-2                      & $96$      & $3 \times  3$         & $1$       & $24 \times  36 \times  96$     \\
$(att1)$ pool3                  & -       & $3 \times  3$         & $2$       & $11 \times  17 \times  96$    \\\hline
conv4-1                      & $128$     & $3 \times  3$            & $1$       & $11 \times  17 \times  128$    \\
conv4-2                      & $128$     & $3 \times  3$            & $1$       & $11 \times  17 \times  128$    \\
$(att2)$ pool4& -       & $3 \times  3$           & 2       & $5 \times  8 \times  128$      \\
\hline
\end{tabular}}
\end{table}

\begin{table}[ht]
\floatconts
  {tab:resnet50}%
  {\caption{ResNet-50 base architecture for classification}}%
  {\renewcommand{\arraystretch}{1.3}\begin{tabular}{r|c|c|c|l|c}
\bfseries Layer    & \bfseries Kernels & \bfseries Kernel Size & \bfseries Strides &                             & \bfseries Output shape     \\ \hline
conv1    & $64$    & $7 \times  7$     & $2$     & \multicolumn{1}{c|}{}       & $224 \times  224 \times  64$ \\ \hline
maxpool  & -       & $3 \times  3$     & $2$     & \multicolumn{1}{c|}{}       & $112 \times  112 \times  64$ \\ \hline
         & $64$    & $1 \times  1$     & $1$     & \multicolumn{1}{c|}{}       &                  \\
conv2\_$\times$   & $64$    & $3 \times  3$     & $1$     & \multicolumn{1}{c|}{($\times  3$)} & $56 \times  56 \times  256$  \\
         & $256$   & $1 \times  1$     & $1$     & \multicolumn{1}{c|}{}       &                  \\ \hline
        & $128$   & $1 \times  1$     & $2$     & \multicolumn{1}{c|}{}       &                  \\
 $(att0)$ conv3\_$\times$     
         & $128$   & $3 \times  3$     & $1$     & \multicolumn{1}{c|}{($\times  4$)} & $28 \times  28 \times  512$  \\
         & $512$   & $1 \times  1$     & $1$     & \multicolumn{1}{c|}{}       &                  \\ \hline
         & $256$   & $1 \times  1$     & $2$     & \multicolumn{1}{c|}{}       &                  \\
$(att1)$ conv4\_$\times$  & $256$   & $3 \times  3$     & $1$     & \multicolumn{1}{c|}{($\times  6$)} & $14 \times  14 \times  512$  \\
         & $1024$  & $1 \times  1$     & $1$     & \multicolumn{1}{c|}{}       &                  \\ \hline
        & $512$   & $1 \times  1$     & $2$     & \multicolumn{1}{c|}{}       &                  \\
$(att2)$ conv5\_$\times$   & $512$   & $3 \times  3$     & $1$     & \multicolumn{1}{c|}{($\times  3$)} & $7 \times  7 \times  512$    \\
         & $2048$  & $1 \times  1$     & $1$     & \multicolumn{1}{c|}{}       &                  \\ \hline
\end{tabular}}
\end{table}

\begin{figure}[h]
\centering
\includegraphics[width=0.69\textwidth]{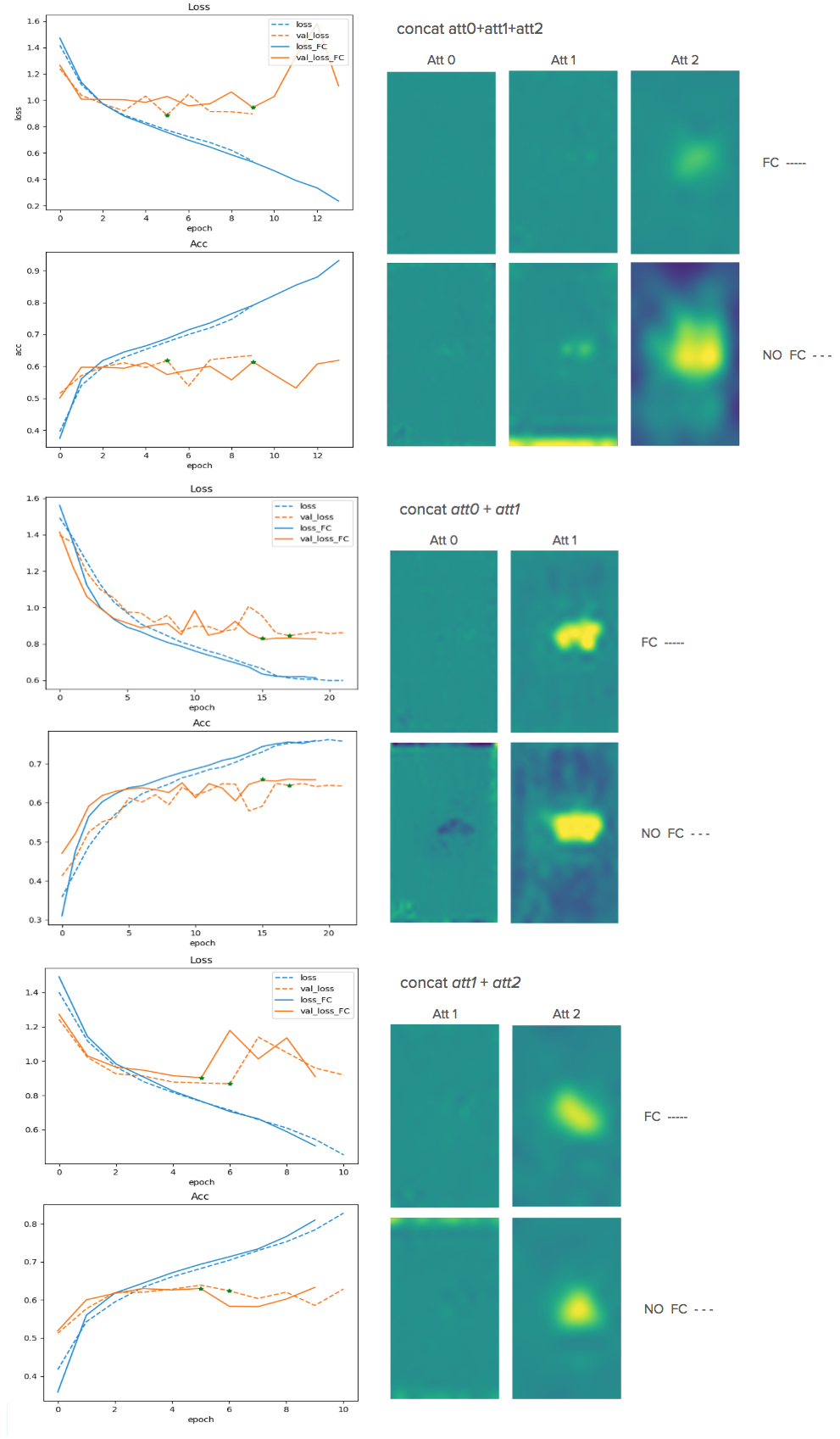}
\caption{Learning curves and visualization for early fusion experiment in VGG-16.}
\label{fig:vgg-16-concat}
\end{figure}

\begin{figure}[h]
\centering
\includegraphics[width=0.71\textwidth]{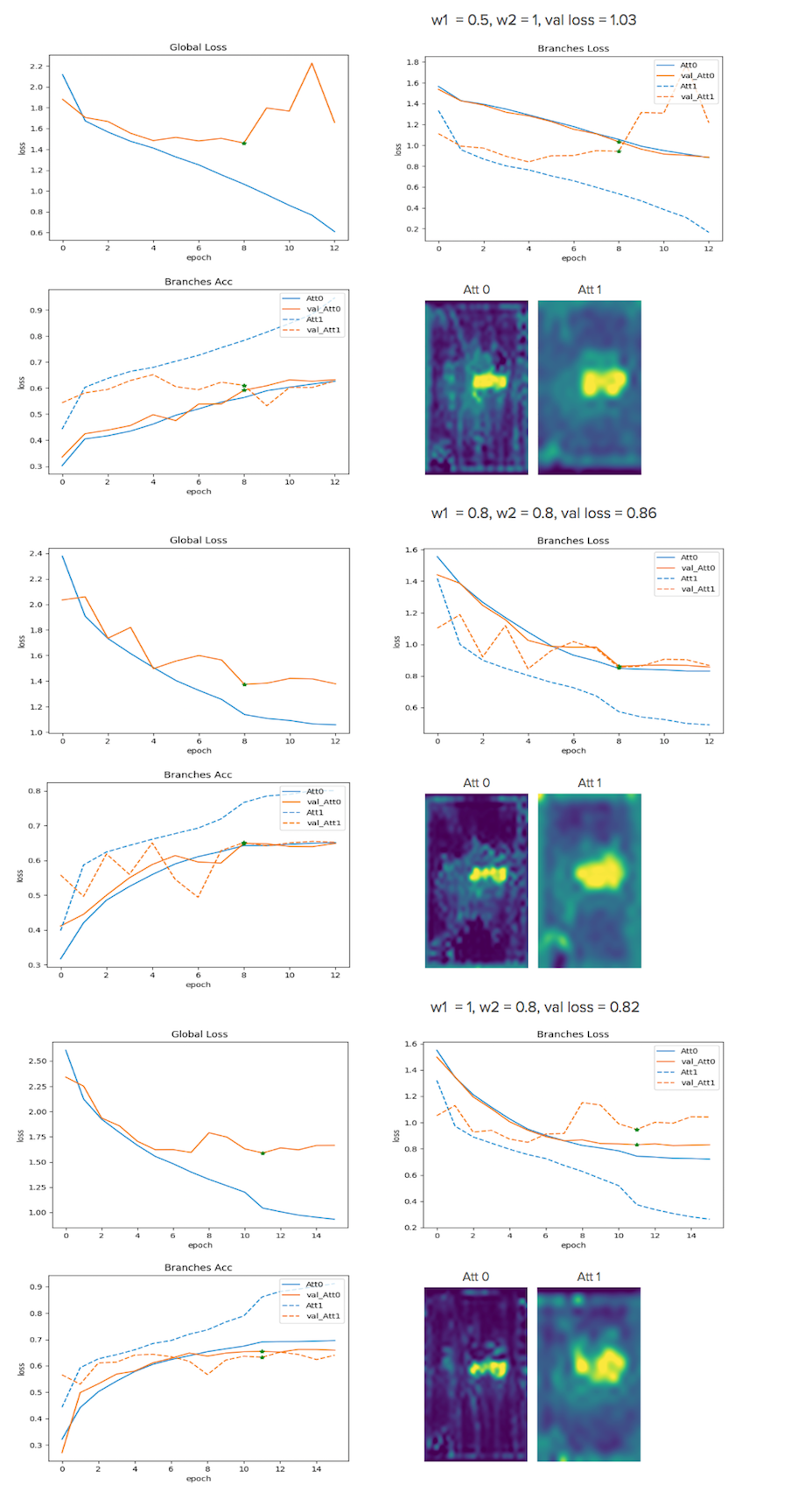}
\caption{Learning curves and visualization for multi-loss experiment in VGG-16.}
\label{fig:vgg-16-multiloss}
\end{figure}

\begin{figure}[h]
\centering
\includegraphics[width=0.8\textwidth]{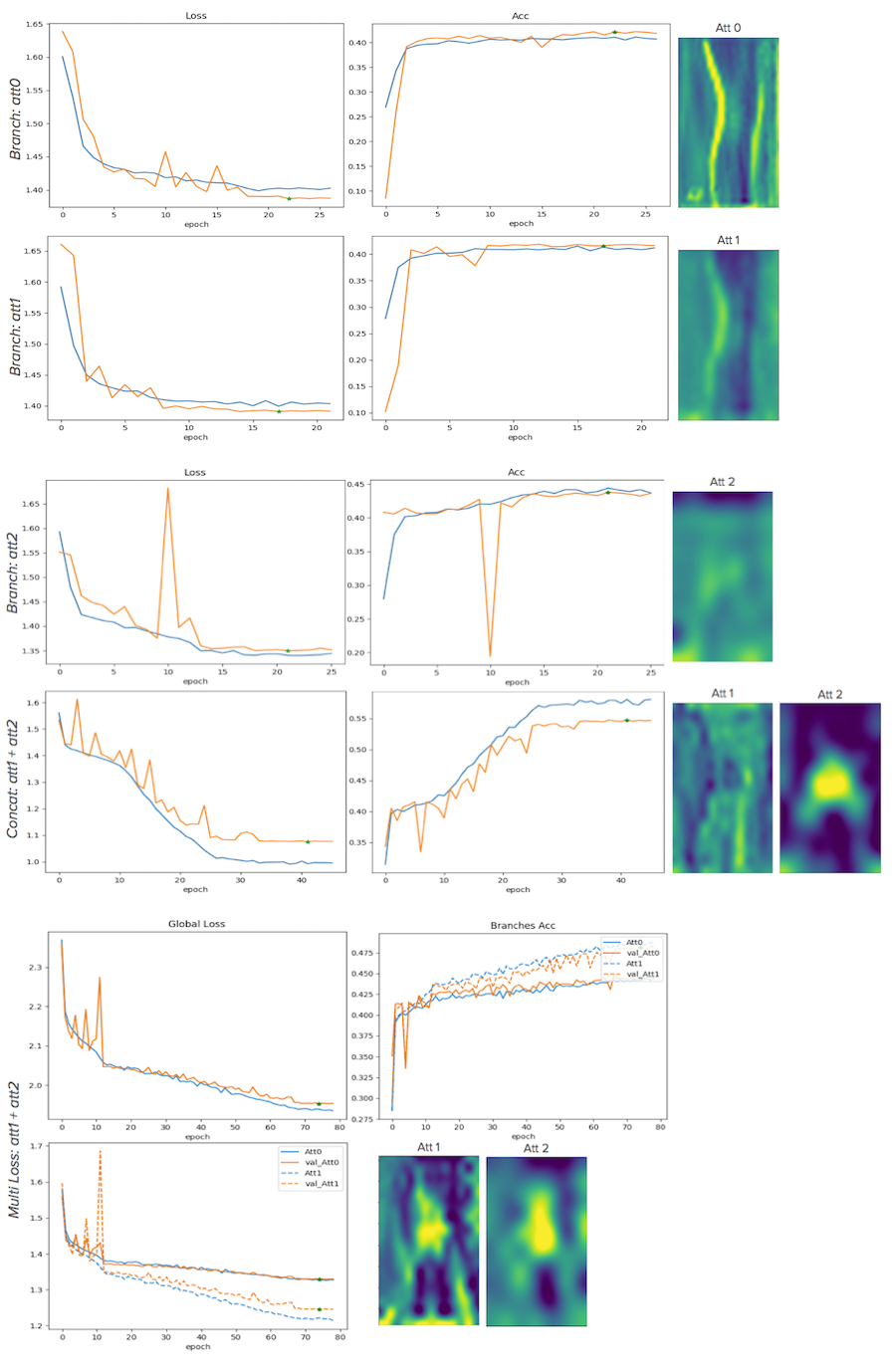}
\caption{Learning curves and visualization for Antony et al. pipeline for classification.}
\label{fig:antony-clsf}
\end{figure}

\begin{figure}[h]
\centering
\includegraphics[width=0.8\textwidth]{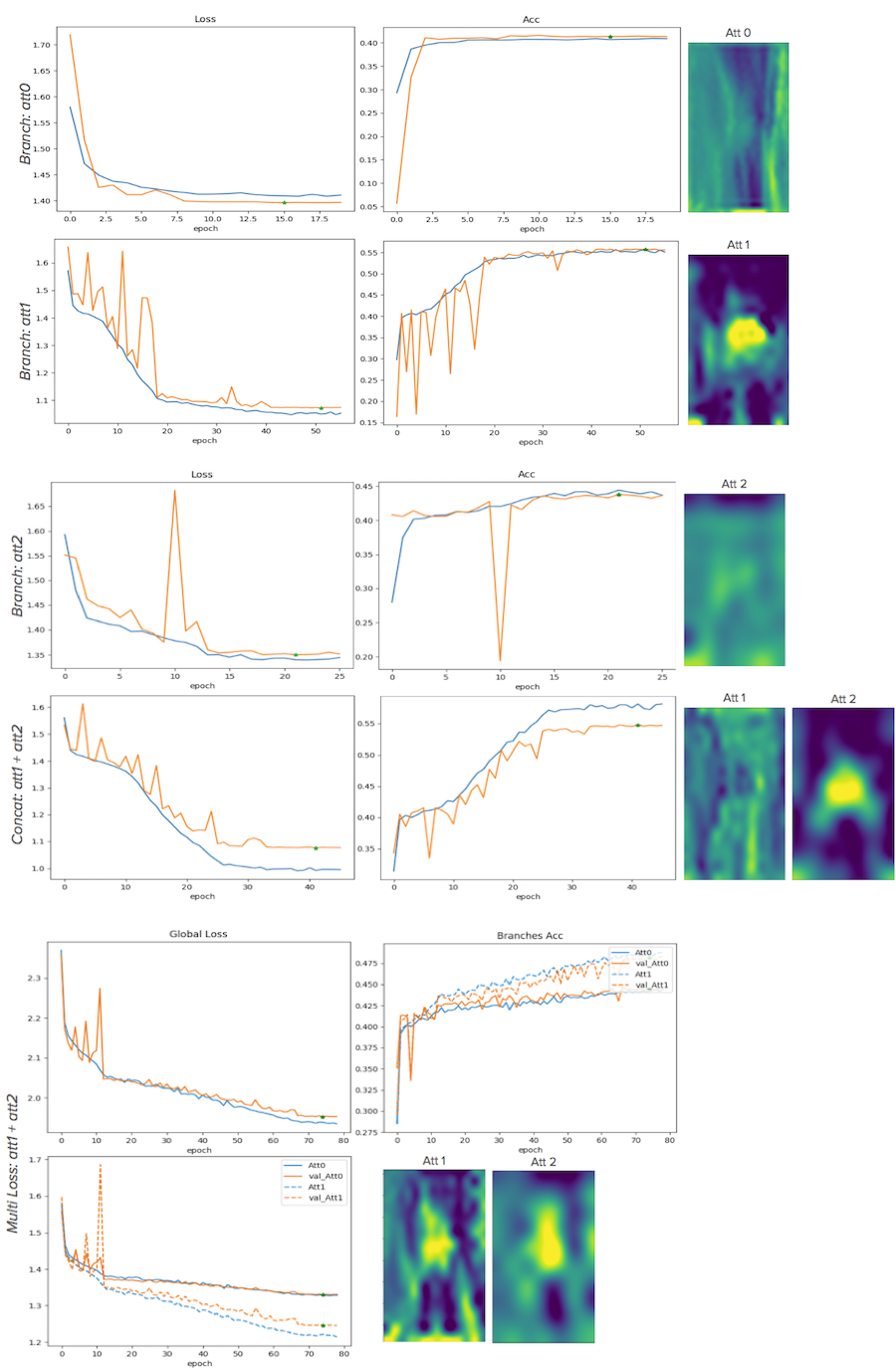}
\caption{Learning curves and visualization for Antony et al. pipeline for jointly classification and regression.}
\label{fig:antony-ext}
\end{figure}

\begin{figure}[h]
\centering
\includegraphics[width=0.8\textwidth]{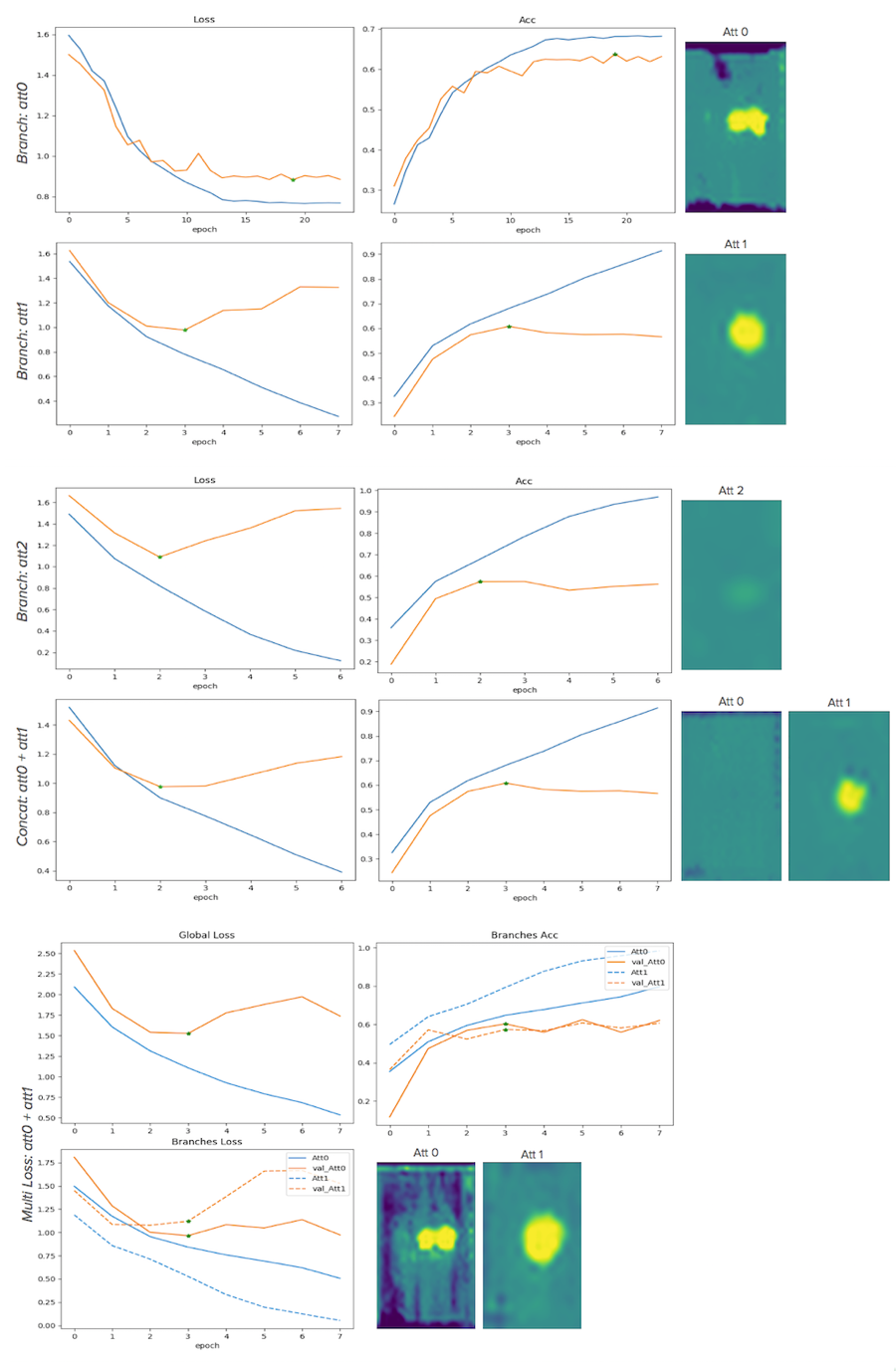}
\caption{Learning curves and visualization for ResNet-50 pipeline.}
\label{fig:resnet-perf}
\end{figure}

\end{document}